\documentclass[conference]{IEEEtran}

\usepackage[usenames, dvipsnames]{color}
\usepackage[hidelinks]{hyperref}
\usepackage{tikz}

\ifCLASSOPTIONcompsoc
    \usepackage[caption=false,font=normalsize,labelfont=sf,textfont=sf]{subfig}
\else
    \usepackage[caption=false,font=footnotesize]{subfig}
\fi

\usepackage{amssymb}

\usepackage{amsthm}
\newtheorem{thm}{Theorem}
\theoremstyle{definition}
\newtheorem{defn}{Definition}

\usepackage{algorithm}
\usepackage{algpseudocode}

\usepackage{amsmath}
\DeclareMathOperator*{\argmin}{\arg\!\min}

\newcommand{\urbanxor}{URBAN\_XOR}
\newcommand{\cP}{{\mathcal{P}}}

\title{Binary Multi-Level Routing Protocol for Mobile Ad Hoc Networks}

\author{
  Anatoliy Zinovyev and Brian L. Mark \\
  email: aszinovyev@gmail.com, bmark@gmu.edu \\
   \\
  George Mason University \\
}

\begin{document}

\maketitle

\begin{abstract}
Routing in mobile ad hoc networks (MANETs) presents a big challenge, especially when support for a large number of nodes is needed. This paper extends the local visibility concept of the recent DHT-based \urbanxor~routing protocol, which aims to reduce routing table sizes while keeping efficiency high. Our main contribution is providing a guarantee that if any two nodes are connected through other nodes, they are able to communicate with each other. We propose a new route acquisition method that aims to reduce the total amount of overhead traffic and improve convergence rate. In addition, we introduce an abstraction for describing the network structure that makes it easy to understand and analyze. Compared to existing approaches in ad hoc routing, the new protocol supports the following features: scalability, guaranteed connectivity assuming network convergence, absence of single points of failure, low path stretch, and mobility.
\end{abstract}

\begin{figure*}[!t]
    \centering
    
    \newcommand{\nodescl}{0.7}
    \newcommand{\edgescl}{1.3}
    \input{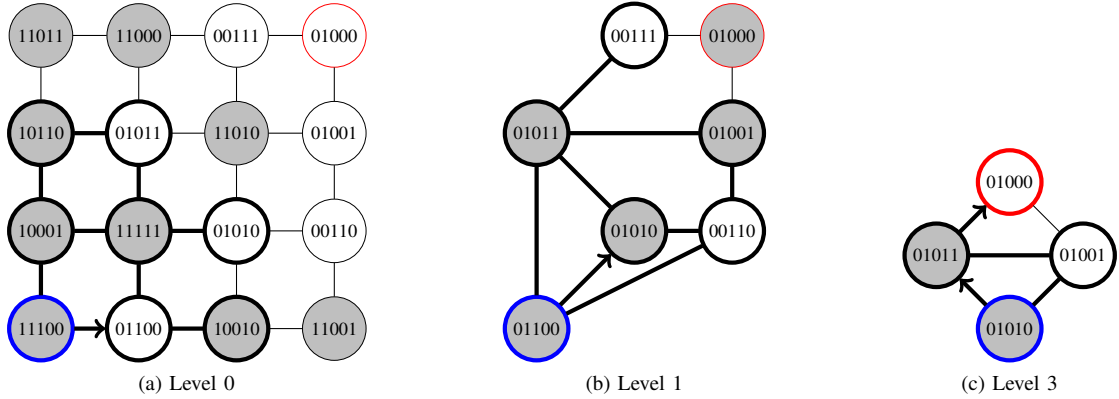}

    \subfloat[Level 0] {
        \begin{tikzpicture}
            \Bn[draw=blue] (n0) at (0,0) {11100};
            \Wn (n1) at (\coordscl{1}, \coordscl{0}) {01100};
            \Bn (n2) at (\coordscl{2}, \coordscl{0}) {10010};
            \bn (n3) at (\coordscl{3}, \coordscl{0}) {11001};
            \Bn (n4) at (\coordscl{0}, \coordscl{1}) {10001};
            \Bn (n5) at (\coordscl{1}, \coordscl{1}) {11111};
            \Wn (n6) at (\coordscl{2}, \coordscl{1}) {01010};
            \wn (n7) at (\coordscl{3}, \coordscl{1}) {00110};
            \Bn (n8) at (\coordscl{0}, \coordscl{2}) {10110};
            \Wn (n9) at (\coordscl{1}, \coordscl{2}) {01011};
            \bn (n10) at (\coordscl{2}, \coordscl{2}) {11010};
            \wn (n11) at (\coordscl{3}, \coordscl{2}) {01001};
            \bn (n12) at (\coordscl{0}, \coordscl{3}) {11011};
            \bn (n13) at (\coordscl{1}, \coordscl{3}) {11000};
            \wn (n14) at (\coordscl{2}, \coordscl{3}) {00111};
            \wn[draw=red] (n15) at (\coordscl{3}, \coordscl{3}) {01000};

            \draw[->,ultra thick]
                (n0) -- (n1);

            \draw[ultra thick]
                (n1) -- (n2)
                (n4) -- (n5)
                (n5) -- (n6)
                (n8) -- (n9)
                (n0) -- (n4)
                (n1) -- (n5)
                (n4) -- (n8)
                (n5) -- (n9)
                ;

            \draw
                (n2) -- (n3)
                (n6) -- (n7)
                (n9) -- (n10)
                (n10) -- (n11)
                (n12) -- (n13)
                (n13) -- (n14)
                (n14) -- (n15)
                (n2) -- (n6)
                (n3) -- (n7)
                (n6) -- (n10)
                (n7) -- (n11)
                (n8) -- (n12)
                (n9) -- (n13)
                (n10) -- (n14)
                (n11) -- (n15)
                ;

        \end{tikzpicture}

        \label{fig:bmlrp-routing-l0}
    }%
    \hfil%
    \subfloat[Level 1] {
        \begin{tikzpicture}
            \Bn[draw=blue] (n1) at (\coordscl{1}, \coordscl{0}) {01100};
            \Bn (n6) at (\coordscl{2}, \coordscl{1}) {01010};
            \Wn (n7) at (\coordscl{3}, \coordscl{1}) {00110};
            \Bn (n9) at (\coordscl{1}, \coordscl{2}) {01011};
            \Bn (n11) at (\coordscl{3}, \coordscl{2}) {01001};
            \Wn (n14) at (\coordscl{2}, \coordscl{3}) {00111};
            \bn[draw=red] (n15) at (\coordscl{3}, \coordscl{3}) {01000};

            \draw[->,ultra thick]
                (n1) -- (n6);

            \draw[ultra thick]
                (n1) -- (n7)
                (n1) -- (n9)
                (n6) -- (n7)
                (n6) -- (n9)
                (n7) -- (n11)
                (n9) -- (n11)
                (n9) -- (n14)
                ;

            \draw
                (n11) -- (n15)
                (n14) -- (n15)
                ;
        \end{tikzpicture}

        \label{fig:bmlrp-routing-l1}
    }%
    \hfil%
    \subfloat[Level 3] {
        \begin{tikzpicture}
            \Bn[draw=blue] (n6) at (\coordscl{0.75}, \coordscl{0}) {01010};
            \Bn (n9) at (\coordscl{0}, \coordscl{0.75}) {01011};
            \Wn (n11) at (\coordscl{1.5}, \coordscl{0.75}) {01001};
            \Wn[draw=red] (n15) at (\coordscl{0.75}, \coordscl{1.5}) {01000};

            \draw[->,ultra thick]
                (n6) -- (n9);
            \draw[->,ultra thick]
                (n9) -- (n15);

            \draw[ultra thick]
                (n6) -- (n11)
                (n9) -- (n11)
                ;

            \draw
                (n11) -- (n15);

        \end{tikzpicture}

        \label{fig:bmlrp-routing-l3}
    }
    
    \caption{Routing in BMLRP}
    \label{fig:bmlrp-routing}
\end{figure*}

\section{Introduction}
\label{sec:introduction}

Over the past thirty years, a number of routing protocols for mobile ad hoc networks (MANETs) have been developed.
In reactive protocols such as AODV~\cite{aodv} and DSR~\cite{dsr}, routes are set up  on demand via route request packets.  Proactive protocols, such as OLSR~\cite{olsr}, maintain routes to destinations by periodically disseminating routing information throughout the network. Reactive protocols generally perform better in mobile scenarios, but can have high latencies and introduce high traffic overhead during route setup.  Hybrid protocols, such as ZRP~\cite{zrp}, attempt to combine the features of the reactive and proactive approaches.  The aforementioned protocols do not scale to networks with large numbers of nodes, e.g., in the thousands or several thousands. 
  
Recently, MANET routing protocols based on routing schemes for distributed hash tables (DHTs) have been proposed.  DHT-based MANET routing protocols  are interesting for several reasons. First is their scalability properties achieved by creating a special network structure and reducing the routing table size. Second is their robustness relative to clustering-based approaches. In clustering-based protocols, nodes are divided into groups, which are subsequently split into smaller groups~\cite{Hong}. Although, such an approach can be scalable to larger networks, critical nodes for controlling clusters~\cite{Hong} and dynamic addresses \cite{Hong,dart,ntk} are used, which may compromise the stability of the network.

DHT-based routing, however, often suffers from the mismatch problem resulting in a high path stretch\footnote{The path stretch is defined as $ \frac{{\rm len}(a,b)}{{\rm dist}(a,b)} $, where ${\rm len}(a,b)$ is the route length between nodes $a$ and $b$ found by the routing protocol and ${\rm dist}(a,b)$ is the length of the shortest path between $a$ and $b$. The length of a path can be measured as either the latency or the number of hops}. In such networks, every node has its own logical identifier (LID) and also stores information about a portion of other nodes with certain LIDs, such that any node should be able to send information to any other node. Because in such networks routing is done on top of the logical structure and the logical addresses do not necessarily correspond to the physical locations, the number of physical hops through which data traverses is often far from optimal \cite{Abid}.

A number of interesting solutions have been proposed for minimizing the path stretch. For example, the Virtual Ring Routing protocol builds a virtual ring where nodes are ordered according to their LIDs~\cite{VRR}. Each node maintains a record of a constant number of nodes with closest addresses and paths to them. When routing data, the next hop with the closest address to the destination's identifier is chosen. The guarantee of constant path stretch relies on the fact that each node knows a total of $O(\sqrt{N})$ nodes (where $N$ is the total number of nodes in the network). Hence, the probability that a node will know a route to the destination is $O(\frac{1}{\sqrt{N}})$, and the expected number of traversed nodes is $O(\sqrt{N})$~\cite{VRR}. Given that the average distance in a wireless ad hoc network is also $O(\sqrt{N})$~\cite{Kleinrock}, the path stretch is constant. Another approach is taken by the 3D~routing protocol, which ensures good path stretch properties by embedding the node LIDs into a 3-dimensional space~\cite{3drp}. Thus, forwarding data is as simple as sending it in the ``right direction.'' Both approaches, however, fail to deal with the network mobility and merging/splitting operations. 

The Binary Multi-Level Routing Protocol (BMLRP) proposed in this paper is closest in approach to the recent KDSR~\cite{kdsr} and \urbanxor~\cite{Pasquini} protocols, which employ a Kademlia DHT~\cite{kademlia} inspired approach for building the network structure. Each node maintains $n$ buckets for storing information about other nodes with address prefixes equal to the node's address prefix. When routing information to any node, the next hop with the longest matching prefix is chosen from among the $n$~buckets. Path efficiency in \urbanxor~is achieved by employing the concept of local visibility, which prioritizes physically close nodes in the routing table. Maintaining such network structure ensures small path stretch, limits protocol overhead and supports mobility better than other routing approaches do~\cite{urbanxor}.

The \urbanxor\ protocol, however, has several drawbacks. The most prominent one is the absence of a guarantee that if two nodes are indirectly connected they are able to communicate \cite{Pasquini}. Another objection is the slow method of acquiring paths to other nodes, which still tends to generate much overhead traffic in mobile scenarios. The proposed BMLRP protocol is a proactive routing protocol that aims to solve both problems altogether.

The rest of the paper is organized as follows. Section~\ref{sec:urbanxor} describes the \urbanxor\ routing protocol in more detail. Section~\ref{sec:bmlrp} introduces BMLRP, which overcomes the major drawbacks of \urbanxor. 
Section~\ref{sec:analysis} discusses routing properties of the new approach. 
Section~\ref{sec:conclusion} concludes the paper.

\section{\urbanxor}
\label{sec:urbanxor}

In the \urbanxor\ protocol each node has a unique permanent $n$-bit identifier that is randomly generated before connecting to the network. The routing table of every node is organized into $n$ buckets, each of size $K$. We will denote $l(a,b)$ as a function of two addresses. The value of $l(a,b)$ is the length of the longest common prefix of $a$ and $b$. For instance, $l(01100, 01000) = 2$ because the biggest common prefix is $01$. When a node $a$ discovers a new neighbor $b$, identifier $b$ is added to the bucket number $l(a,b)$ of node $a$.

After the node connects to its direct (both physical and virtual) neighbors, it starts filling the $n$ buckets by asking for missing nodes. When a new node is discovered, an abstract virtual link is built to it. Thus, multi-hop routes in the physical space are stored in a reduced form. At the same time nodes passively overhear the traffic to find new routes without loading the network. The local visibility concept ensures that the impact of any update in the network is limited by close nodes.

Similarly to Kademlia DHT \cite{kademlia}, when node $a$ is routing data to $b$, the first node selects an address $c$ from its routing table such that $l(c,b) > l(a,b)$. Because each next node that forwards the data has longer common prefix with the destination, the data eventually arrives. For this to be always true, each node in the network must have at least one entry in each bucket if an appropriate identifier exists. The \urbanxor\ protocol, however, might maintain empty buckets and thus fail to satisfy this property~\cite{Pasquini}.

\section{The Proposed Protocol:  BMLRP}
\label{sec:bmlrp}

Consider a node with a unique permanent $n$-bit address $a$, similarly to \urbanxor. We say that all nodes $b$ in the network form a level-$i$ network with respect to $a$ if $l(a,b) \ge i$. We also refer to level-$i$ nodes as the nodes from the level-$i$ network with respect to $a$.  The level-$i$ routing table of $a$ denotes a graph containing all visible level-$i$ nodes of $a$. For instance, the level-0 network will contain all nodes, which is also defined as the \emph{physical network}. The level-3 network will contain all nodes that have the first 3 bits equal to the first 3 bits of $a$. Because we assume all identifiers are distributed evenly, the number of level-3 nodes will be approximately 8 times smaller than the physical network size. As a general rule, the level-$i$ network contains approximately $\frac{N}{2^i}$ nodes, where $N$ is the size of the physical network.

After a node joins the physical network, it starts creating virtual links to the nearby level-1 nodes. In a sense, it connects to them the same way it connects to its physical neighbors. We now abstractly consider a level-1 network similarly to a level-0 network according
to all of its properties. The idea is to ascend levels up to the point where the number of visible level-$m$ nodes is equal to the total number of all level-$m$ nodes. 

\begin{figure*}[!t]
    \centering

    \newcommand{\nodescl}{1}
    \newcommand{\edgescl}{1}
    \input{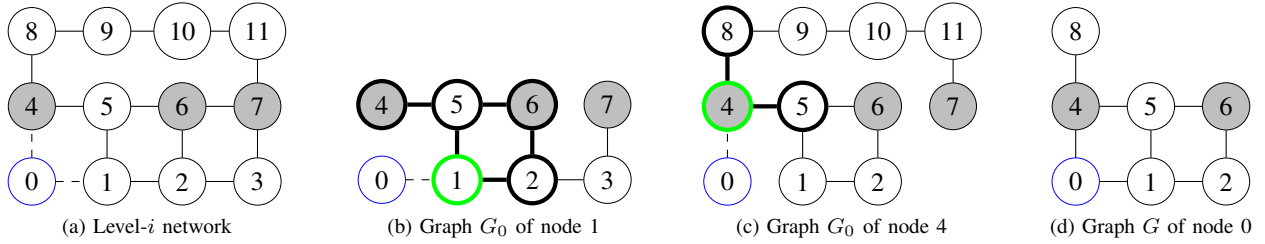}

    \subfloat[Level-$i$ network] {
        \begin{tikzpicture}
            \wn[draw=blue] (n0) at (\coordscl{0}, \coordscl{0}) {0};
            \wn (n1) at (\coordscl{1}, \coordscl{0}) {1};
            \wn (n2) at (\coordscl{2}, \coordscl{0}) {2};
            \wn (n3) at (\coordscl{3}, \coordscl{0}) {3};
            \bn (n4) at (\coordscl{0}, \coordscl{1}) {4};
            \wn (n5) at (\coordscl{1}, \coordscl{1}) {5};
            \bn (n6) at (\coordscl{2}, \coordscl{1}) {6};
            \bn (n7) at (\coordscl{3}, \coordscl{1}) {7};
            \wn (n8) at (\coordscl{0}, \coordscl{2}) {8};
            \wn (n9) at (\coordscl{1}, \coordscl{2}) {9};
            \wn (n10) at (\coordscl{2}, \coordscl{2}) {10};
            \wn (n11) at (\coordscl{3}, \coordscl{2}) {11};

            \draw[dashed]
                (n0) -- (n1)
                (n0) -- (n4)
                ;

            \draw
                (n1) -- (n2)
                (n2) -- (n3)
                (n2) -- (n6)
                (n6) -- (n5)
                (n7) -- (n6)
                (n4) -- (n8)
                (n11) -- (n7)
                (n8) -- (n9)
                (n9) -- (n10)
                (n10) -- (n11)
                (n1) -- (n5)
                (n3) -- (n7)
                (n4) -- (n5)
                ;
        \end{tikzpicture}
        
        \label{fig:bmlrp-routes-all}
    }%
    \hfil%
    \subfloat[Graph $G_0$ of node 1] {
        \begin{tikzpicture}
            \wn[draw=blue] (n0) at (\coordscl{0}, \coordscl{0}) {0};
            \Wn[draw=green] (n1) at (\coordscl{1}, \coordscl{0}) {1};
            \Wn (n2) at (\coordscl{2}, \coordscl{0}) {2};
            \wn (n3) at (\coordscl{3}, \coordscl{0}) {3};
            \Bn (n4) at (\coordscl{0}, \coordscl{1}) {4};
            \Wn (n5) at (\coordscl{1}, \coordscl{1}) {5};
            \Bn (n6) at (\coordscl{2}, \coordscl{1}) {6};
            \bn (n7) at (\coordscl{3}, \coordscl{1}) {7};
            
            \draw[dashed]
                (n0) -- (n1);

            \draw[ultra thick]
                (n1) -- (n2)
                (n4) -- (n5)
                (n1) -- (n5)
                (n2) -- (n6)
                (n5) -- (n6)
                ;

            \draw
                (n2) -- (n3)
                (n3) -- (n7)
                ;                
        \end{tikzpicture}

        \label{fig:bmlrp-routes-nb1}
    }%
    \hfil%
    \subfloat[Graph $G_0$ of node 4] {
        \begin{tikzpicture}
            \wn[draw=blue] (n0) at (\coordscl{0}, \coordscl{0}) {0};
            \wn (n1) at (\coordscl{1}, \coordscl{0}) {1};
            \wn (n2) at (\coordscl{2}, \coordscl{0}) {2};
            \Bn[draw=green] (n4) at (\coordscl{0}, \coordscl{1}) {4};
            \Wn (n5) at (\coordscl{1}, \coordscl{1}) {5};
            \bn (n6) at (\coordscl{2}, \coordscl{1}) {6};
            \bn (n7) at (\coordscl{3}, \coordscl{1}) {7};
            \Wn (n8) at (\coordscl{0}, \coordscl{2}) {8};
            \wn (n9) at (\coordscl{1}, \coordscl{2}) {9};
            \wn (n10) at (\coordscl{2}, \coordscl{2}) {10};
            \wn (n11) at (\coordscl{3}, \coordscl{2}) {11};

            \draw[dashed]
                (n0) -- (n4);

            \draw[ultra thick]
                (n4) -- (n5)
                (n4) -- (n8)
                ;

            \draw
                (n1) -- (n2)
                (n1) -- (n5)
                (n2) -- (n6)
                (n5) -- (n6)
                (n7) -- (n11)
                (n8) -- (n9)
                (n9) -- (n10)
                (n10) -- (n11)
                ;
        \end{tikzpicture}

        \label{fig:bmlrp-routes-nb2}
    }%
    \hfil%
    \subfloat[Graph $G$ of node 0] {
        \begin{tikzpicture}
            \wn[draw=blue] (n0) at (\coordscl{0}, \coordscl{0}) {0};
            \wn (n1) at (\coordscl{1}, \coordscl{0}) {1};
            \wn (n2) at (\coordscl{2}, \coordscl{0}) {2};
            \bn (n4) at (\coordscl{0}, \coordscl{1}) {4};
            \wn (n5) at (\coordscl{1}, \coordscl{1}) {5};
            \bn (n6) at (\coordscl{2}, \coordscl{1}) {6};
            \wn (n8) at (\coordscl{0}, \coordscl{2}) {8};

            \draw
                (n0) -- (n1)
                (n1) -- (n2)
                (n0) -- (n4)
                (n1) -- (n5)
                (n2) -- (n6)
                (n4) -- (n5)
                (n5) -- (n6)
                (n4) -- (n8)
                ;
        \end{tikzpicture}

        \label{fig:bmlrp-routes-visible}
    }

    \caption{Routes propagation}
    \label{fig:bmlrp-routes}
\end{figure*}

When node $a$ is routing data to node $d$, it will choose the closest next hop $b$ such that $l(b,d) > l(a,d)$. In other words, the first bit that is not equal between $a$ and $d$ must be equal between $b$ and $d$. Similarly to Kademlia DHT and \urbanxor, each new node routing traffic has a longer common prefix with $d$, thus avoiding loops. To communicate with same-level nodes, source routing is used. Assuming $a$ sees $b$ in its level-$i$ routing table, it prepends path $c_1$--$c_2$--...--$c_k$--$b$ ($c_j$ are level-$i$ nodes) to the packet and sends the data to $c_1$ which in turn will forward it according to the given path. Sending the packet to $c_1$ involves routing on an underlying level-$j$ network ($j < i$). Similarly, a new path of level-$j$ nodes is prepended, as well as, paths for all underlying networks. Given that nodes' addresses are distributed uniformly, the number of levels and the length of the source routing prefix is bounded by the logarithm of the total number of nodes. A real implementation will not include full $n$-bit addresses in the source routing prefix, but the details are omitted here.

For simplicity, from now on we will label a level-$i$ node white if the $(i+1)$-th bit of its address is 0; otherwise the node is black. Figure~\ref{fig:bmlrp-routing} shows a simplified example of routing data from 11100 to 01000, with subfigures demonstrating different levels of the network with respect to the destination. The blue nodes denote the routing nodes; the red node on each level indicates the destination. Bold entries indicate nodes and links known by the routing node. The journey starts on the physical level depicted by Figure~\ref{fig:bmlrp-routing-l0}. The source and destination nodes differ in the first bit, therefore 11100 transmits the data to a close physical white node -- 01100. Figure~\ref{fig:bmlrp-routing-l1} shows the level-1 network with respect to the destination, where the routing node is now 01100. Node 01100 chooses the next hop $b$ such that $l(b, 01000) > l(01100, 01000) = 2$; $b$ equals 01010. When node 01010 receives the message, it will find the destination address in its level-3 routing table 
and will route the data through an intermediate level-3 node.

\begin{defn}
    We say that nodes $a$ and $b$ are connected in level-$i$ network if $a$ and $b$ send information about close level-$i$ nodes and edges to each other, including those returned by Algorithm~\ref{alg:routes}.
    
    \label{defn:connected}
\end{defn}

For each edge sent by node $a$ to its connected level-$i$ neighbor $b$, a sequence of level-$i$ nodes $c_1$--$c_2$--$c_3$--...--$c_k$ must be included, where $c_1$ and $c_2$ (or $c_1$ and $a$ if there is no $c_2$) are the nodes incident to the edge and $c_2$--$c_3$--...-$c_k$--$a$--$b$ is the path through which the edge was transmitted from $c_2$ to $b$. Node $a$ must choose the shortest such path $c_1$--$c_2$--...--$c_k$ and ensure that no $c_j$ equals $b$. Records of the paths contain shortcuts from a previously defined dictionary instead of full $n$-bit identifiers.

When node $b$ has just connected to $a$ in level-$i$ network, $a$ will send a subset of close level-$i$ nodes and edges to $b$. For any update in $a$'s level-$i$ routing table, $a$ will calculate a new subset of close nodes and edges and send an update to $b$ -- the difference between the new subset and the old subset. This way, the overhead is kept low and, unlike in \urbanxor, all necessary information is propagated reactively, making convergence fast. Each node combines graphs received from its level-$i$ neighbors and forms a new graph $G$, also called the level-$i$ routing table.

We will now examine which nodes and edges must be propagated by each node in order to satisfy the connectivity requirement. Assume all nodes in Algorithm~\ref{alg:routes} are level-$i$ nodes.

\begin{algorithm}[H]
    \caption{Include necessary routes}
    \label{alg:routes}
    
    \begin{itemize} 
        \item   For each connected neighbor $b$, node $a$ constructs a graph $G_b$ by merging the graphs received from all of its neighbors except $b$ and adding all direct edges to these neighbors, and adding the edge $a$--$b$. Node $a$ then calculates a subset $G_{outb}$ of $G_b$ defined as follows.

        \item   Consider every path $b$--$a$--$c_1$--$c_2$--...--$c_k$--$d$ in $G_b$; call this path $\cP$ (all nodes in $\cP$ are different). Without loss of generality, let $a$ be white. All nodes and edges of $\cP$ will be included in $G_{outb}$ if all of the following hold:
        \begin{enumerate}
            \item   $d$ is black , 
            \item   $c_j$ are white ($1 \le j \le k$) ,
            \item   for any $c_j$ there is no black node $f$ in $G_b$, such that the distance
                    from $f$ to $c_j$ is smaller than the distance 
                    from $c_j$ to $d$ in $\cP$ and the distance from $c_j$ to $b$ in $\cP$.
        \end{enumerate}

        \item   Exclude node $b$ from $G_{outb}$.
    \end{itemize}
\end{algorithm}

Figure~\ref{fig:bmlrp-routes} shows an example in which nodes only propagate edges returned by Algorithm~\ref{alg:routes}. Node~0 connects to nodes~1 and 4. By Algorithm~\ref{alg:routes}, node 1's graph $G_0$ is depicted in Figure~\ref{fig:bmlrp-routes-nb1}. Paths that will be included in $G_{out0}$ are 0--1--5--4, 0--1--5--6, 0--1--2--6. Path 0--1--2--3--7 will not be included because the distance from 6 to 2 is shorter than both paths 0--1--2 and 2--3--7. Figure~\ref{fig:bmlrp-routes-nb2} shows a similar graph of node 4. Here, the only 2 paths satisfying the requirement are 0--4--5 and 0--4--8. Node~0 combines the information received from nodes 1 and 4, and forms a routing table $G$ depicted in Figure~\ref{fig:bmlrp-routes-visible}.

To ensure connectivity between any two nodes in the network, it is sufficient to make every same-color node in level-$i$ graph $G$ a neighbor on level~$(i+1)$. It can be foreseen, however, that this will lead to big routing tables on higher levels. Therefore, a more restrictive approach is needed. Algorithm~\ref{alg:connect} characterizes which nodes are necessary to connect to in order to satisfy the connectivity requirement. Node $a$ indicates the current node. The graph $G$ is assumed to be a level-$i$ routing table of $a$.
The procedure  connect$(x)$ indicates that $x$ will become a level-$(i+1)$ neighbor of $a$. Function bit$(x, l)$ returns the $(i + l)$-th bit of node $x$'s address, assuming the first bit is numbered~0, and XOR$(c, d)$ denotes bitwise exclusive OR.

Without loss of generality, assume $a$ is black in level-$i$ network.

\begin{figure*}[!t]
    \centering

    \newcommand{\nodescl}{1}
    \newcommand{\edgescl}{1}
    \input{graphs.tex}

    \begin{tikzpicture}
        \wn (n0) at (\coordscl{0}, \coordscl{0}) {0};
        \bn (n1) at (\coordscl{1}, \coordscl{1}) {1};
        \bn (n2) at (\coordscl{1}, \coordscl{-1}) {2};
        \bn (n3) at (\coordscl{-1}, \coordscl{-1}) {3};
        \bn (n4) at (\coordscl{-1}, \coordscl{1}) {4};
        \wn (n5) at (\coordscl{-2}, \coordscl{0}) {5};
        \bn (n6) at (\coordscl{-3}, \coordscl{1}) {6};
        \bn (n7) at (\coordscl{-3}, \coordscl{-1}) {7};
        \wn (n8) at (\coordscl{-4}, \coordscl{0}) {8};
        \bn (n9) at (\coordscl{-5}, \coordscl{1}) {9};
        \bn (n10) at (\coordscl{-5}, \coordscl{-1}) {10};
        \wn (n11) at (\coordscl{2}, \coordscl{0}) {11};
        \wn (n12) at (\coordscl{4}, \coordscl{0}) {12};
        \bn (n13) at (\coordscl{5}, \coordscl{1}) {13};
        \bn (n14) at (\coordscl{5}, \coordscl{-1}) {14};
        \bn (n15) at (\coordscl{6}, \coordscl{0}) {15};

        \draw[dashed]
            (n1) -- (n2)
            (n3) -- (n4)
            (n4) -- (n1)
            (n4) -- (n6)
            (n6) -- (n7)
            (n7) -- (n10)
            (n9) -- (n10)
            (n2) -- (n14)
            (n14) -- (n15)
            (n15) -- (n13)
            ;

        \draw
            (n0) -- (n1)
            (n0) -- (n2)
            (n0) -- (n3)
            (n0) -- (n4)
            (n0) -- (n5)
            (n5) -- (n6)
            (n5) -- (n7)
            (n5) -- (n8)
            (n8) -- (n9)
            (n8) -- (n10)
            (n0) -- (n1)
            (n0) -- (n2)
            (n0) -- (n11)
            (n11) -- (n12)
            (n12) -- (n13)
            (n12) -- (n14)
            (n12) -- (n15)
            ;
    \end{tikzpicture}
        
    \caption{Minimal connecting}
    \label{fig:bmlrp-connect}
\end{figure*}

\begin{algorithm}[H]
    \caption{Connect node $a$ to necessary nodes}
    \label{alg:connect}

    \begin{itemize} 
        \item   connect$(b)$ for all black level-$i$ neighbors $b$ of $a$.

        \item   For every white level-$i$ neighbor $b$ of $a$: 
                let $s$ be the set of all black neighbors of $b$, execute ConnectInside$(s, 0)$.

        \item   Consider every path $\cP = $ $a$--$b$--$c_1$--$c_2$--...--$c_k$--$d$ in $G$.  \\
        connect($d$) if all of the following hold:
        \begin{enumerate}
            \item   $d$ is black ,
            \item   $b$ and $c_j$ ($1 \le j \le k$) are white ,
            \item   $c_j$ ($1 \le j < k$) do not have black neighbors ,
            \item   for any $c_j$ ($1 \le j \le k$) there is no black node $f$ in $G$, such that the distance
                    from $f$ to $c_j$ is smaller than the distance
                    from $c_j$ to $d$ in $\cP$ and the distance from $c_j$ to $a$ in $\cP$ ,
            \item   choose a black neighbor $e_1$ of $b$ and a black neighbor $e_2$ of $c_k$,
                    such that XOR$(e_1, e_2)$ is minimized.
                    $e_1$ equals $a$
        \end{enumerate}
    \end{itemize} 
\end{algorithm}

\begin{algorithm}
    \begin{algorithmic}[1]
        \Procedure{ConnectInside}{${\rm nodes}, l$}
            \State ${\rm white} \gets \{x \in {\rm nodes}: {\rm bit}(x, l) = 0\}$
            \State ${\rm black} \gets \{x \in {\rm nodes}: {\rm bit}(x, l) = 1\}$
            
            \If {${\rm white} = \emptyset$ \textbf{or} ${\rm black} = \emptyset$}
                \State \textbf{return}
            \EndIf

            \State $w,b \gets \argmin_{x \in {\rm white}, y \in {\rm black}}{\rm XOR}(x, y)$

            \If {${\rm bit}(a, l) = 0$}
                \If {$w = a$}
                    \State connect($b$)
                \EndIf
                
                \State ConnectInside(${\rm white}$, $l + 1$)
            \Else
                \If {$b = a$}
                    \State connect($w$)
                \EndIf

                \State ConnectInside(${\rm black}$, $l + 1$)
            \EndIf
        \EndProcedure
    \end{algorithmic}
\end{algorithm}

Figure~\ref{fig:bmlrp-connect} illustrates Algorithm~\ref{alg:connect} (solid edges show level-$i$ links, dashed edges indicate level-$(i+1)$ connections). First, note how black neighbors of white nodes are organized into trees. Additionally, some black nodes are chosen to connect to distant black nodes. For instance, nodes~4 and 6 are connected, and nodes~2 and 14 are connected.

So far we have only discussed which nodes are necessary to connect to. A real implementation will, however, try to maintain a constant number of closest neighbors on each level. This assumption will be necessary for 
the analysis in Section~\ref{sec:analysis:stretch}.

\section{Analysis}
\label{sec:analysis}

\subsection{Absence of Stale Routes}
\label{sec:analysis:staleroutes}

To guarantee the connectivity between any two nodes, assuming network convergence, we need to be certain that the protocol will not send data through a non-existing edge. Suppose, node $c_k$ has learned about edge $c_1$--$c_2$ from node $c_{k-1}$. As explained earlier, a record of this edge contains path $\cP = $ $c_1$--$c_2$--...--$c_{k-2}$ through which the edge has been sent to $c_{k-1}$ and then to $c_k$. In case some $c_j$ and $c_{j+1}$ ($1 \le j < k$) disconnect, $c_{j+1}$ eliminates a record of the edge, and if $j+1 < k$, sends an update to $c_{j+2}$. Eventually, $c_k$ receives an update after which it removes the record of edge $c_1$--$c_2$ containing path $\cP$.

Hence for every record of an edge, the path recorded will not contain any broken edges. When any two nodes disconnect, this event will be propagated through every such path, and the edge will be removed from the memory of all nodes.

\subsection{Connectivity}
\label{sec:analysis:connectivity}

Theorems~\ref{thm:connectivity-levels} and \ref{thm:connectivity-colors} establish 
two important connectivity properties maintained by the proposed protocol.

\begin{thm}
    Let a level-$i$ network be a connected graph. Then, the level-$(i+1)$ network is also a connected graph.

    \begin{IEEEproof}
        \normalfont Assume, the level-$(i+1)$ graph is not connected (1). Now, let $c_0$ and $d$ be nodes from two distinct graph components in level-$(i+1)$ graph, such that the distance from $c_0$ to $d$ in level-$i$ network is minimized (2). Also, without loss of generality assume $c_0$ and $d$ are black on level~$i$. If $c_0$ and $d$ are neighbors on level~$i$, by Algorithm~\ref{alg:connect} they are connected on level~$(i+1)$ and this contradicts (1). Otherwise, let $\cP = $ $c_0$--$c_1$--...--$c_k$--$d$ be a shortest path from $c_0$ to $d$ in level-$i$ network. By the assumption (2), nodes $c_j$ ($1 \le j \le k$) are white.

        Assume, node $c_0$ does not know some edge in $\cP$. Then, by Algorithm~\ref{alg:routes} some node $c_j$ ($1 \le j \le k$) does not forward this information to $c_{j-1}$ because it knows such black node $f$ that the distance from $f$ to $c_t$ (for some $j < t \le k$) is shorter than the distance from $d$ to $c_t$ in $\cP$ and the distance from $c_t$ to $c_{j-1}$ in $\cP$. If $c_t$ is not connected to either $c_0$ or $d$ on level~$(i+1)$, then this must contradict the minimality defined in (2). Otherwise, if $c_t$ is connected to both $c_0$ and $d$ on level~$(i+1)$, $c_0$ and $d$ must be from the same graph component in level-$(i+1)$ graph and this also contradicts (2). Hence, $c_0$ knows the whole $\cP$. 
        
        Similarly, $c_0$ also knows edges from $c_{k-1}$ to all black neighbors of $c_{k-1}$, $d$ knows $\cP$, $d$ knows edges from $c_1$ to all black neighbors of $c_1$. By Algorithm~\ref{alg:connect} some black neighbor of $c_1$ must be connected to some black neighbor of $d$. However, all black neighbors of $c_1$ form a connected level-$(i+1)$ graph and all black neighbors of $d$ form a connected level-$(i+1)$ graph. Hence, $c_0$ and $d$ are in the same graph component in the level-$(i+1)$ network.
    \end{IEEEproof}

    \label{thm:connectivity-levels}
\end{thm}

\begin{thm}
    Let a level-$i$ network be a connected graph. Then, every node in this network must know at least one level-$i$ node of the opposite color, if such exists.

    \begin{IEEEproof}
        \normalfont Suppose, node $a$ does not know any node of the opposite color. Without loss of generality, let $a$ be black. Now, let $d$ be a white node, such that the distance from $a$ to $d$ in level-$i$ network is minimized. By this assumption, all nodes $c_j$ along a shortest path $a$--$c_1$--$c_2$--...--$c_k$--$d$ must be black. Similarly to the previous proof, $a$ must know a path to $d$, which contradicts the original assumption.
    \end{IEEEproof}

    \label{thm:connectivity-colors}
\end{thm}

Given the discussion in Section~\ref{sec:analysis:staleroutes} and Theorems~\ref{thm:connectivity-levels}~and~\ref{thm:connectivity-colors}, we are able to say that a node $a$ will always be able to forward a data packet with destination $d$ to the next hop $b$, such that $l(b,d) > l(a,d)$, satisfying the connectivity requirement.

\subsection{Path stretch}
\label{sec:analysis:stretch}

We will now roughly estimate the path stretch of the proposed protocol. Each data packet is routed on (possibly not all) level-0, level-1, ..., level-$k$ networks with respect to the destination until it finally reaches the destination. We assume equal properties across all levels 0, ..., $k - 1$, including each node maintaining a constant number of closest neighbors on each level.

The path stretch when routing data over a level-$(i+1)$ network has to be some constant multiple $q$ ($q \ge 1$) of the stretch when routing data over a level-$i$ network. Because $C_0\sqrt{N}$ is the average distance in a wireless network of $N$ identical uniformly placed nodes~\cite{Kleinrock} and level-$i$ nodes only route data to close level-$i$ nodes with longer common address prefix with destination, the distance in the physical network between the two nodes must be proportional to $\sqrt{2^i}$. We also know that the maximum level $k$ over which data is routed is less than $\log_2N$. We now find the average number of traversed hops:
\begin{equation*}
\begin{split}
    {\rm hops} & = C \left (1 + \sqrt{2} q + \sqrt{2^2} q^2  + ... + \sqrt{2^{k-1}} q^{k-1} + \sqrt{N} q^k \right ) \\
              & < C \left( (\sqrt{2} q)^{k-1} \sum_{i = 0}^{\infty} (\sqrt{2}q)^{-i} + \sqrt{N} q^k \right ) \\
               & < C \left ( \sqrt{N} q^k \frac{\sqrt{2} q}{\sqrt{2}q - 1} + \sqrt{N} q^k \right )  \\
               & < C \left ( \frac{2\sqrt{2}q - 1}{\sqrt{2}q - 1} \right ) \sqrt{N} q^{\log_2 N}  \\
               & = C \left ( \frac{2\sqrt{2}q - 1}{\sqrt{2}q - 1} \right ) N^{\frac{1}{2} + \log_2 q} .
\end{split}
\end{equation*}
Thus, an upper bound for the average number of hops is given by
\begin{equation*}
    O({\rm hops}) = O \left ( N^{\frac{1}{2} + \log_2 q} \right ) .
\end{equation*}
Dividing by $\sqrt{N}$, we find the upper bound for the average path stretch:
\begin{equation*}
    O({\rm stretch}) = O \left ( N^{\log_2 q} \right ) .
\end{equation*}
For example, when each new level increases the path stretch by 10\%, we have $O({\rm stretch}) \approx O \left ( \sqrt[7]{N} \right )$ and $O({\rm hops}) \approx O \left ( N^{0.64} \right )$, which can be considered scalable in many situations. 

Communication between level-$i$ neighbors might be done over underlying level-$(i-1)$, level-$(i-2)$, level-$(i-3)$, ... networks, depending on the number of visible nodes on each level. Hence, increasing the routing tables decreases the constant $q$, and the upper bounds for average number of hops and stretch also decrease asymptotically.

\subsection{Average node degree}
\label{sec:analysis:nodedegree}

Despite several advantages, BMLRP does not guarantee that the routing table sizes for higher level networks remain low. To analyze this aspect, we implemented\footnote{\url{https://github.com/aszinovyev/bmlrp-simple}} and tested Algorithms~\ref{alg:routes} and \ref{alg:connect} on a static network to track the average degree of nodes across each level.

\begin{figure}[H]
   \includegraphics[width=0.95\linewidth]{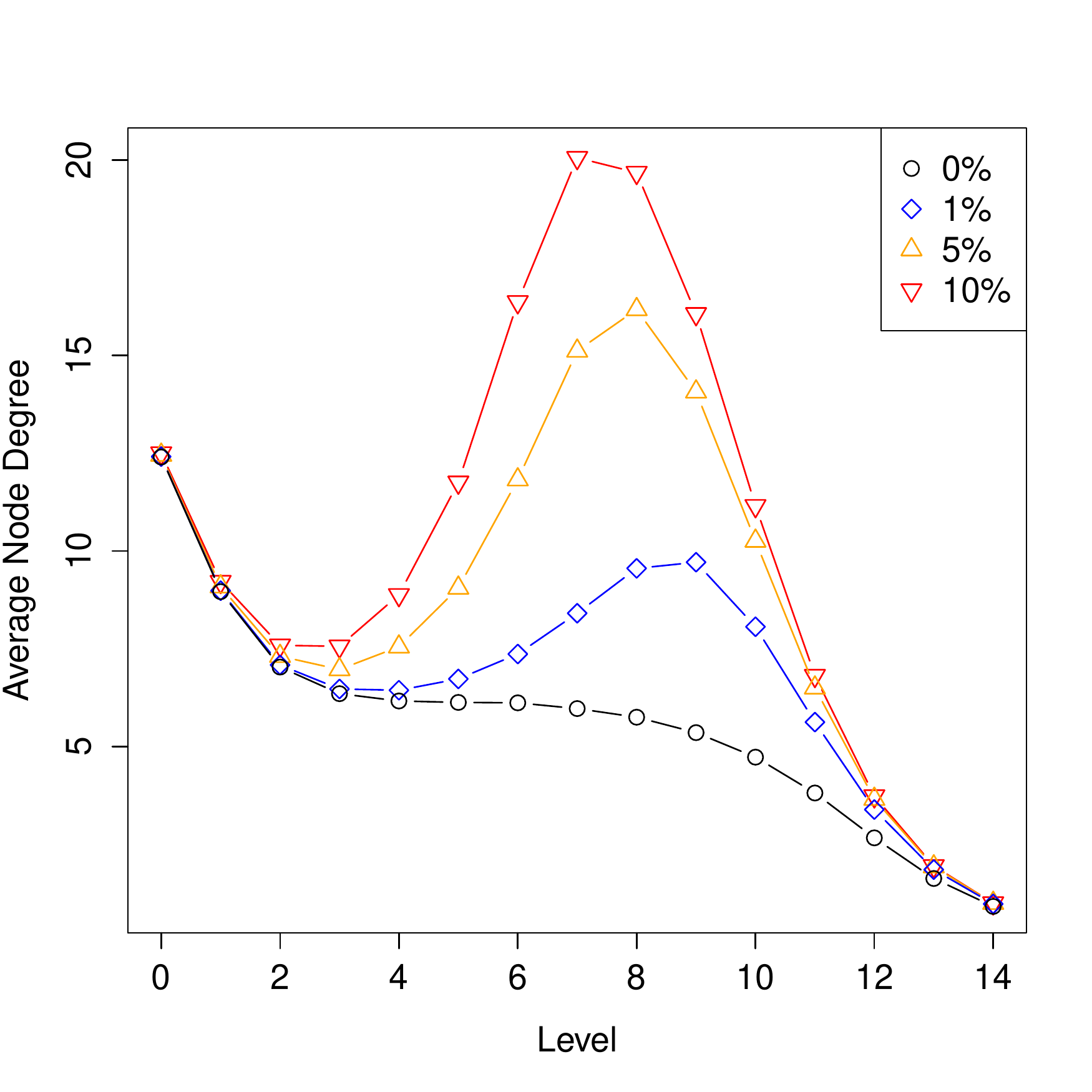}

   \caption{Average node degree vs.\ network level~$i$.}
    \label{fig:degrees}
\end{figure}

In Fig.~\ref{fig:degrees}, the black line shows the simulation result on $2^{14}$ identical nodes uniformly distributed in a square area. Note that the average number of neighbors first stabilizes around the value~6 and then continues decreasing as the network level $i$ grows. Additionally, we randomly connected 1\%, 5\% and 10\% nodes in the network independently of their coordinates. The blue, orange and red lines show that when random long-range edges are added, the higher level networks become close world networks and the number of mandatory connections fails to decrease. The impact, however, is small; the red line demonstrates that even when 10\% of the nodes are randomly connected, the average number of level-$i$ neighbors in the network does not exceed 21 for $2^{14}$ nodes.

\section{Conclusion}
\label{sec:conclusion}

We have developed BMLRP, a new proactive DHT-based routing protocol for MANETs, which has several important features in comparison to previous protocols: scalability, guaranteed connectivity assuming network convergence,  absence of single points of failure, low path stretch, and mobility. The proposed BMLRP protocol has a similar setup to the \urbanxor\ protocol~\cite{Pasquini}, which employs a Kademlia DHT~\cite{kademlia} inspired approach for building the network structure. Unlike the \urbanxor\ protocol, however, BMLRP guarantees that if two nodes are indirectly connected they are able to communicate. Furthermore, the proposed protocol generates lower overhead traffic and converges faster than \urbanxor\ in mobile scenarios.  

Connectivity properties of the proposed protocol were proven, and growth rates of the path stretch and number of hops for routes were given. The average node degree as a function of the network level $i$ was studied numerically. In ongoing work, we are implementing the protocol using ns-3 to evaluate path~stretch, overhead, delivery ratio and other properties through simulation.

\bibliography{bmlrp}
\bibliographystyle{unsrt}

\end{document}